\documentclass[twocolumn,showpacs,preprintnumbers,amsmath,amssymb]{revtex4}

\usepackage{graphicx}
\usepackage{dcolumn}
\usepackage{bm}

\begin{document}


\title{Maximum Entropy Principle, Equal Probability {\em a Priori}
and Gibbs Paradox}

\author{Hao Ge$^{1,2}$}
\email{gehao@fudan.edu.cn}
\author{Hong Qian$^3$}%
\email{qian@amath.washington.edu}
\affiliation{$^1$School of Mathematical Sciences and
Centre for Computational Systems Biology, Fudan University, Shanghai 200433, PRC.
$^2$Department of Chemistry and Chemical Biology, Harvard
University, Cambridge, MA 02138, USA.
$^3$Department of Applied
Mathematics, University of Washington, Seattle, WA 98195, USA.}

\date{\today}

\begin{abstract}

We show that the information-theoretic maximum entropy (MaxEnt)
approach to deriving the canonical ensemble theory is mathematically
equivalent to the classic approach of Boltzmann, Gibbs and
Darwin-Fowler.  The two approaches, however, ``interpret'' a same
mathematical theorem differently; most notably observing mean-energy
in the former and energy conservation in the latter. However,
applying the same MaxEnt method to the grand canonical ensemble
fails, while the correct statistics is obtained if one
carefully follows the classic approach based on Boltzmann's
microcanonical {\em equal probability a priori}. One does not need
to invoke quantum mechanics, and there is no Gibbs paradox. MaxEnt
and related minimum relative entropy principle are based on the
mathematical theorem concerning large deviations of rare
fluctuations. As a scientific method, it requires classical
mechanics or other assumptions to provide meaningful {\em
prior distributions} for the expected-value based statistical
inference. A naive assumption of uniform prior is not valid in
statistical mechanics.
\end{abstract}

\maketitle

There exist several frameworks, based on information theory and/or
statistical inferences
\cite{Jaynes57,Szilard,Mandelbrot62}, which have been put
forward as possible alternatives to the Boltzmann-Gibbsian
foundation of statistical thermodynamics \cite{Boltz64,Gibbs}. At
the same time, there has been a growing body of ``statistical
mechanical'' studies of systems and processes with no thermal
molecular origin, ranging from signal processing to combinatorial
optimization, to neural networks \cite{neuronet}. Almost all these
studies claim Shannon's information theory as their foundation
\cite{Cover91}.  These approaches are used to deduce probability
distributions for fluctuating quantities which usually has no
connection to Newtonian mechanics.

This paper reexamines these approaches while clarifying and
contrasting the differences between the classic approach to
statistical mechanics and the new ones. The principle of maximum
entropy (MaxEnt), or minimum Kullbak-Leibler relative entropy
\cite{hobson}, which is at the heart of these information-based
approach, was first proposed by E.T. Jaynes as an alternative
foundation for statistical mechanics \cite{Jaynes57}. On the other
hand, in information theory, following the collective work of many,
it is now generally accepted that the minimum relative entropy
principle is a mathematical theorem
\cite{shore80,Cover81,Banavar10}. It is also known as Gibbs
conditioning to probabilists \cite{gibbs_con}. Consider a sequence
of identical, independently distributed (i.i.d.) discrete random
variables $X_1, X_2, \cdots, X_n$, with the state space
$\mathcal{S}=\{\omega_1,\omega_2,...,\omega_M\}$ ($M$ could be
infinite) and prior distribution $\Pr\{X_i=\omega_m\}=p_m$. (In
classical mechanics, the $\omega_i$ are called microstates.) The
theorem states
\begin{equation}
    \lim_{n\rightarrow\infty}
    \Pr\left\{ X_1 = \omega_m\Big| \sum_{i=1}^n
    h(X_i) = n\alpha\right\}
    = p^*_m,
\label{eq_1}
\end{equation}
where $h(\cdot)$ is a function defined on the state space
$\mathcal{S}$ and $p^*_m$ has the ``canonical'' form of
$e^{\lambda_0+\lambda_1 h(\omega_m)}\cdot p_m$. Two constants
$\lambda_0$ and $\lambda_1$ are determined according to
\begin{equation}
    \sum_{m=1}^{M} p^*_m = 1, \ \ \
    \sum_{m=1}^{M} h(\omega_m)p^*_m = \alpha.
\label{2_constraints}
\end{equation}
The first condition enforces normalization; the second one is
interpreted as ``conditioned on observing the mean value for
$h(\cdot)$''. We shall call $p_m$ the {\em prior}. The form of
$p^*_m$ is exactly the same as that derived through minimizing the
relative entropy
\begin{equation}
    H\left(\left\{p^*_m\right\}||\left\{p_m\right\}\right)
        = \sum_{m=1}^{\infty} p^*_m\ln\left(\frac{p^*_m}{p_m}\right),
\end{equation}
under the constraints given in (\ref{2_constraints}). It is
important to point out that MaxEnt is really the minimum relative
entropy with uniform prior.

This theorem is concerned with the conditional distribution of a
collection of individual samples, given that some quantity averaged
over the large number of individual samples shows highly unlikely
behavior.  Note that if the observed sample mean is the expected
value of the
prior, then $\lambda_0=\lambda_1=0$. This theorem is closely related
to the large deviation theory of empirical distribution (i.e.,
histogram)
\begin{equation}
    L_n(\omega_m)=\frac{1}{n}\sum_{i=1}^n
        \delta_{\{X_i=\omega_m\}},
\end{equation}
where the delta function $\delta_{\{X_i=\omega_m\}}=1$ if
$X_i=\omega_m$; and zero otherwise. According to the strong law of
large number in probability theory, we know that this empirical
distribution converges to the prior distribution $\{p_m\}$ of $X_i$.
Moreover, the level-2 Sanov theorem in large deviation theory
\cite{Touchette09,gibbs_con} states that for any
set $\mathcal{C}$ of
probability distribution, we have
\begin{equation}
  \lim_{n\rightarrow \infty}\frac{1}{n}\ln\Pr\{L_n\in \mathcal{C}
            \}=-\inf_{\mu\in \mathcal{C}}\{I(\mu)\},\label{eq_5}
\end{equation}
where $I(\mu)=\sum_{i=1}^n
\mu(\omega_i)\ln\left[\mu(\omega_i)/p_i\right]$ is the relative
entropy of $\mu$ with respect to the prior distribution $p_i$.
Therefore, the relative entropy can be interpreted as the ``free
energy'' of deviation in the sense of a distribution \cite{Qian01}.
And at this juncture, the two free energies, one from the theory 
of large deviation and one in the theory of Markov dynamics \cite{ep_fd_10},
agree.

If we set $\mathcal{C}=\{\mu:\langle h(\cdot)\rangle_{\mu}=\alpha\}$
as the space of probability distribution with given constraints, then
for any give distribution $\mu$,
\begin{equation}
  \Pr\left\{L_n=\mu\Big|\frac{1}{n}\sum_{i=1}^nh(X_i) = \alpha\right\}
            =\Pr\{L_n=\mu|\mathcal{C}\}\rightarrow 0,
\end{equation}
unless $\mu=\mu^*$ where $\mu^*$ satisfies $I(\mu^*)=\inf_{\mu\in
\mathcal{C}}\{I(\mu)\}$, i.e. with minimum relative entropy. It is
implied that the empirical distribution $L_n$ is dominated by
$\mu^*$ when $n\rightarrow\infty$. Furthermore, since the
distributions of different $X_i$ under the constraint
$\frac{1}{n}\sum_{i=1}^nh(X_i) = \alpha$ are identical, the limiting
distribution $\mu^*$ for $L_n$ also holds for each $X_i$.

If one assumes uniform prior distribution in the canonical ensemble due
to ignorance and constrains based on the observed mean energy $h(\cdot)$, then
the posterior distribution $h(X_i)$ is just the exponential,
canonical distribution. On the other hand, Jaynes \cite{Jaynes57}
argued that the entropy of statistical mechanics and the entropy in
information theory are principally the same thing, and simply
maximizing the entropy
\begin{equation}
    S\left(\left\{p^*_m\right\}\right)
        = \sum_{m=1}^{\infty} p^*_m\ln p^*_m
\end{equation}
under some constraint on the mean observations would give the correct
canonical distribution. Hence such an optimizing argument is
mathematically equivalent to the previous theorem, and consequently
statistical mechanics could be re-interpreted as a particular
application of a general theory of logical inference and information
theory \cite{ben-naim}. While Jaynes' approach to statistical
mechanics, as well as the widely-used minimum relative entropy
principle in information theory, is based on observations of
mean-energy, the classic approach of Boltzmann, Gibbs and
Darwin-Fowler to statistical mechanics interprets the same theorem
differently.

For the canonical ensemble, suppose it is a part of a larger
microcanonical ensemble consisting of $N$ closed, identical
canonical ensembles. Let $X_i$ represent the microstate of the
$i$-th canonical ensemble, say momenta and positions $(p, q)$. Then
the high dimensional vector $X=(X_1,X_2,...,X_N)$ represents a
microstate in the microcannonical ensemble. Let the function
$e(X_i)$ be the energy of the $i$-th canonical ensemble, and the
total energy $E_{tot}=\sum_{i=1}^N e(X_i)$.  The law of classical
mechanical energy conservation says the $X$ is only confined in the
subspace $\{E_{tot}=H\}$ where $H$ is the given total energy of the
larger microcanonical ensemble.  The notion of equal a priori
probability further assumes that the probability of $X$ is equally
distributed on such a subspace \cite{footnote2}.  The marginal
distribution of each $X_i$ is then exponentially dependent on
$e(X_i)$ when $N$ tends to infinity. Boltzmann's most probable state
method and Darwin-Fowler's steepest descent method are all based on
such a setup and are mathematically equivalent \cite{Ponczek76}.
Note that Boltzmann, Gibbs and Darwin-Fowler deal with the
convergence of empirical distributions as in Eq. (\ref{eq_5}) rather
than marginal distribution as in the theorem (\ref{eq_1}). However,
when $N$ tends to infinity, the limiting empirical distribution is
the same as the limiting marginal distribution.

The distribution for the high-dimensional microstate $X$ in
Boltzmann's approach, subjected to the energy conservation
$E_{tot}=H$, is exactly the same as that of the MaxEnt approach
conditioned on observing $\{\overline{e}=H/N\}$. Hence they are
mathematically equivalent. However, subtle differences exist in
their interpretations: For the MaxEnt approach, one must first
assume the existence of a prior distribution for the canonical ensemble
even without a constraint on the mean energy. In the
classic approach, the equal a priori probability of the entire
microcanonical ensemble can be verified from such a uniform prior
distribution of the independent subsystems without any constraint in the
MaxEnt principle.  That is why Jaynes called this framework
``subjective thermodynamics'' \cite{Jaynes57}.

However, the reasnoning behind using a uniform prior distribution as
the most suitable one when one knows nothing of any random variable
is only empirical, and one must be very careful when applying it to
a specific scientific problem. For example, if one only knows the
mean particle number in a grand-canonical ensemble, this principle
would conclude that the particle number distribution is likewise
exponential (i.e., geometric). But the experimentally observed
distribution is Poissonian when the particle is nearly independent,
whether distinguishable or not (see below for detailed discussion).
Hence only justifying the form of the energy distribution in the
canonical ensemble is not a sufficient proof for the validity of the
MaxEnt principle as substitute for the classical statistic
mechanics.  In other words, the maximum entropy or minimum relative
entropy principle, by itself, can never tell you the prior
distribution. The prior distribution has to be supplied by the
specific problem to which the principle is applied. Of course, for
the purpose of data analysis exclusively, this technique could be
quite useful in supplying a minimal model maximizing the degree of
freedoms beyond the given constraints \cite{Bialek}. Professional
statisticians would also use other methods to test the uniform prior
hypothesis after the analysis of the data.

Now the central question arises: What are exactly the prior
distributions of energy and particle number for the grand-canonical
ensemble? Gibbs tried to answer this question more than one hundred
years ago, starting from the equal probability priori. His
derivation for energy fluctuation was highly successful, but for the
grand-canonical ensemble with fluctuating particle numbers, a
difficulty known as Gibbs paradox arises: Whether or not the phase
space volume $\phi(E,v,n)=\int d^{3n}q\ d^{3n}p$ used in
grand-canonical ensemble should be divided by $n!$. It is now
understood that for microcanonical or canonical ensembles, both with
fixed particle number, the paradox is not a well-defined problem
\cite{Ray1984Baker67}.

Similar to the deviation for canonical ensembles, and still suppose
a large microcanonical ensemble with total energy, volume and
particle number invariant. The box further consists of $N$ open,
identical small grand-canonical ensembles each with fixed volume $
v$ and mean particle number $\langle n \rangle$. They are
statistically identical but not rigorously independent. The
phase-space uniformity states that the high-dimensional microstate
space consists of all the $N\langle n \rangle$ particles in the
large box uniformly distributed in {\em position and momentum}
\cite{footnote2}, and ask what is the distribution of particle
numbers within a small grand-canonical ensemble. Hence the natural
methodology is to calculate the number of high-dimensional
microstates corresponding to a given energy $E$ and particle numbers
$n$ in a grand-canonical volume. This number of high-dimensional
microstates would give the weight (probability) of such a microstate
in the smaller subsystem. The relation between a small subsystem and
the rest of the ``reservoir'' is a rather subtle issue, which has
been repeatedly emphasized in statistical physics.

Textbooks \cite{Mc} often proceed in the same manner as in the
treatment of the canonical ensemble through Boltzmann's most
probable distribution method. This is a little misleading. The key
to this problem lies on how to go about reconstructing the
high-dimensional microstate from those low-dimensional microstates
for each subsystem. There is no problem for the canonical ensemble,
since one can obtain the high-dimensional microstate simply from
linking all the microstates of each subsystem together. However, in
the case of distinguishable particles, we must take into account the
partition of all $N\langle n\rangle$ particles into the $N$
identical subsystems for the grand canonical ensemble. Let $m_i$ be
the number of grand canonical ensembles whose microstates containing
$n_i$ particles with energy $E_i$. Hence, for any possible
distribution $\{m_i\}$ of the microstates in grand canonical
ensembles, there are two kinds of partitions: one is a partition of
these occupation numbers $\{m_i\}$ into a total of $N$ subsystems;
the other is the partition of all the $N\langle n\rangle$ particles
(i.e. labeling particles) into the possible set $\{n_i\}$. The
canonical ensemble only deals with the former, and in textbooks, for
grand canonical ensembles, they also only consider the former one,
while the factorial $n!$ comes from the latter partition of
particles \cite{Ray1984Baker67,footnote1}.

Hence, the number of all high-dimensional microstates
corresponding to $\{m_i\}$ is given by

$$W(\{m_i\})=\frac{N!}{\sum_i m_i!}\times \frac{(N\langle n\rangle)!}{\sum_i \left (n_i!\right )^{m_i}},$$
subject to the three constrains:
\begin{eqnarray}
\sum_i m_i=N,\nonumber\\
\sum_i E_im_i=E_{tot},\nonumber\\
\sum_i n_im_i=N\langle n\rangle,\nonumber
\end{eqnarray}
which could be maximized to derive the correct statistics of grand
canonical ensembles.

For indistinguishable particles, the weight for each
high-dimensional microstate in the large microcanonical ensemble is
already different from the distinguishable case and the factorial
naturally arises due consideration of the phase space volume. Hence
it is well-known that although the $n!$ would not appear because of
the partition of particles into small subsystems, it would emerge
from the phase space volume in this case. Therefore, Gibbs paradox
is definitely not related to quantum mechanics, and the partition
function for grand-canonical ensemble should be written as
\begin{equation}
    \Xi=\sum_{E,n}\frac{Q(E,v,n)}{n!}e^{-\mu n},
\end{equation}
where $Q(E,v,n)$ is the partition function for the canonical ensemble.

For independent distinguishable particles, one could understand the
$n!$ from another perspective. Due to the phase space uniformity
assumption, the position distribution for each particle is uniformly
distributed in this large system with total volume $Nv$. Then at a
certain time, the probability for each particle belonging to a
specific subsystem is $\frac{1}{N}$. Notice that the total number of
particle is $N\langle n\rangle$, hence the distribution of the
particle number in this subsystem is Binomial with parameter
$(N\langle n\rangle, \frac{1}{N})$. When $\langle n\rangle$ is fixed
and $N$ tends to infinity, it converges to a Poisson distribution
with mean $\langle n\rangle$. The factorial just comes out from the
expression of the Poisson distribution
\begin{equation}
             p_n=\frac{\lambda^n}{n!}e^{-\lambda},
\end{equation}
where $\lambda=\langle n\rangle$. This is known as Poisson
statistics for a point process, which has been experimentally
verified in number fluctuation measurement based on fluorescence
correlation spectroscopy (FCS) \cite{qian90}. Furthermore, when $N$
tends to infinity, the positions of the particles must converge to
the well-known Poisson point process and the number of particles
within a certain space is just its counting process.

Let us now come back to the maximum entropy or minimum relative
entropy principle.  It is worth noticing that the phase space
uniformity is of course another form of maximum entropy for the
microcanonical ensemble without any additional constraint
\cite{footnote2} but is different from Jaynes' framework. There is
even confusionregarding the fact that the derivation of the canonical ensemble
distribution by Darwin-Fowler is an application of maximum entropy
approach. This is not the case. Althouth they are based on same
mathematical theorem, they are definitely different interpretations.
What Darwin-Fowler did was to derive the distribution of the
subsystem from the whole phase space uniformity assumption
\cite{Ponczek76}. They did not mention anything like the uniform
prior distribution of the subsystem. The most important element in
Darwin-Fowler's interpretation is still the role of {\em
conservation of energy} at the level of a whole, isolated system,
the First Law of Thermodynamics. They actually justified a special
version of the law of large number in the empirical distribution
space for canonical ensemble, and finally got the limiting
distribution which was exactly Boltzmann's most probable state
\cite{Ponczek76}. We clarify a confusion regarding their terminology. The
``mean" in their work is just the mean occupation number of each
microstate of the subsystem, which is exactly the probability rather
than the real mean of the fluctuating energy.

Jaynes' information approach to classical mechanics is a method of
statistical inference based on macroscopic observables, i.e.,
expected values, in contrast to main stream statistics whose
inferences are often based on samples. In both approaches, a prior
in the absence of any measurement can only be subjective. In the
present paper, we have shown that the Principle of Maximum Entropy
can not fully replace the classical Boltzmann-Gibbs statistical
mechanics precisely because the latter built their ``prior'' based
on (1) uniformity in Newtonian mechanical phase space, and (2)
conservation of energy, number, etc. These two assumptions are
fundamentally outside any logical inference approach.  The case in
point is the grand canonical ensemble: mechanical phase space
uniformity necessarily leads to a Poisson distribution as the prior
for the number distribution of independent classical particles in an
infinitesimal open box.

L. Szilard and B. Mandelbrot also advanced another line of
interpretation for classical thermodynamics, called {\em purely
phenomenological theory}, based on the theory of sufficient
statistics \cite{Szilard,Mandelbrot62}. Interestingly, it is also
based on the above mentioned mathematical theorem, and it asserts
that all the macroscopic thermodynamic quantities are exactly the
sufficient statistics of their microscopic fluctuations. The theory
gives the correct distribution when a given ensemble has been
perturbed but the new system still has the conservation law. It
implies that all of the distributions in statistical mechanics must belong
to the exponential family of probability distributions
\cite{Jeffreys60}.

In the present study, we clarified E.T. Jaynes' MaxEnt approach to
the statistical thermodynamic based on information theory, and its
relation to classical statistical mechanics. It is found that
correctly determining a prior distribution is the central issue,
which could not be addressed in general from only information theory
or statistical inference.  Of course, as a mathematical theory, the
theorem of minimum relative entropy could be applied everywhere and
not just be confined to mechanics or physics. It justifies the
diverse use of ``statistical mechanics'', and explains why it works
as a fundamental tool in information theory.  More importantly, the
mathematical theorem also tells us that the concepts of entropy and
relative entropy are both mathematical constructions, both of which
naturally arise in the asymptotic probability of large deviations
\cite{Touchette09}.

It is arguable that information theory, at least in its mathematical
presentation, is a statistical theory endowed with the concept of
entropy.  This perspective naturally resolves a nagging issue that
troubles ``information'' as a more general theoretic concept: The
relation between information and knowledge \cite{gleick_11}. It
is well understood that thermodynamics is about what is impossible
(for macroscopic systems) and what is very unlikely. It provides
{\em constraints} on molecular processes, but it cannot specify
their mechanisms.  Knowledge is ultimately in the mechanism.  There
seems to be a contradistinction between ``statistics'' and
``knowledge.''

There is another, dynamic origin of the
concept of entropy and relative entropy (or free energy). It has
been shown recently that they are emergent properties of any
Markovian processes \cite{ep_fd_10}. The original Shannon's information theory for coding, however, is a static one. 

We thank Ken Dill, Jin Feng,  Michael Fisher, Chris Jarzynski, Steve
Presse, Jin Wang and Ziqing Zhao for stimulating discussions. H. Ge
acknowledges support by NSFC 10901040 and
specialized Research Fund for the Doctoral Program of Higher
Education (New Teachers) 20090071120003. H. Ge thanks 
Prof. X.S. Xie and members of his group for hospitality 
and support.


\begin{thebibliography}{99}
\bibitem{footnote2}
The rigorous definition of equal a priori probability and entropy of a
microcanonical ensemble for continuous cases are riddled with mathematical
subtleties. We refer the reader to Khinchin's book \cite{khinchin} for details.
\bibitem{footnote1}
The history and different approaches to resolving Gibbs paradox is
controversial.  A more detailed account of the subject will be
published elsewhere.
\end{thebibliography}
\end{document}